\documentclass[prd,preprint,showpacs]{revtex4}
\usepackage{amsmath}
\usepackage{latexsym}
\usepackage{amsfonts}
\usepackage{graphicx}
\usepackage{bm}

\begin{document}

\title{Observational constraint on dynamical evolution of dark energy}

\author{Yungui Gong}
\email{gongyg@cqupt.edu.cn}
\affiliation{College of Mathematics and Physics, Chongqing
University of Posts and Telecommunications, Chongqing 400065, China}

\author{Rong-Gen Cai}
\email{cairg@itp.ac.cn}
\affiliation{Institute of Theoretical Physics, Chinese Academy of
Sciences, Beijing 100190, P. R. China} \affiliation{College of
Mathematics and Physics, Chongqing University of Posts and
Telecommunications, Chongqing 400065, China}
\author{Yun Chen, Zong-Hong Zhu}
\email{zhuzh@bnu.edu.cn}
\affiliation{Department of Astronomy, Beijing Normal University,
Beijing 100875, China}

\begin{abstract}
We use the Constitution supernova, the baryon acoustic oscillation,
the cosmic microwave background, and the Hubble parameter data to
analyze the evolution property of dark energy. We obtain different
results when we fit different baryon acoustic oscillation data combined with
the Constitution supernova data to the Chevallier-Polarski-Linder model.
We find that the difference stems from the different values of $\Omega_{m0}$.
We also fit the observational data to the model independent piecewise constant parametrization.
Four redshift bins with boundaries at $z=0.22$, $0.53$, $0.85$ and $1.8$ were chosen
for the piecewise constant parametrization of the equation of state
parameter $w(z)$ of dark energy. We find no significant evidence for
evolving $w(z)$. With the addition of the Hubble parameter, the
constraint on the equation of state parameter at high redshift is
improved by 70\%. The marginalization of the nuisance parameter connected to the supernova
distance modulus is discussed.

\end{abstract}


\preprint{0909.0596}

\maketitle

\section{Introduction}

Since the discovery of the late time cosmic acceleration by the Type
Ia supernova (SnIa) observations \cite{acc1,acc2}, a lot of efforts
have been made to understand the driving force behind the cosmic
acceleration. The standard models in cosmology and particle physics
give no answer to this problem. To address the problem, one needs to
modify either the left hand side or the right hand side of Einstein
equation.  Modifying the left hand side means that general
relativity is modified, models such as the Dvali-Gabadadze-Porrati
model \cite{dgp}, and $f(R)$ gravity
\cite{fr1,fr2,fr3,fr4,fr5,fr6,fr7,fr8,fr9,fr10}, have been proposed
along this line of reasoning. On the other hand, in the framework of Einstein gravity, an exotic form of
matter with negative pressure, dubbed as dark energy, has to be
introduced into the right hand side of Einstein equation to explain
the phenomenon of cosmic acceleration. However, the nature and
origin of dark energy remain a mystery. Many parametric and
nonparametric model-independent methods were proposed to study the
property of dark energy, see for example
\cite{astier01,huterer,par2,par3,alam,jbp,par4,par1,
par5,gong04,gong05,gong06,goliath,gong08,Ishak2007,rev7,sahni,chris,lu08,cpl1,linder03}
and references therein.

Recently, it was claimed that the flat $\Lambda$CDM model is
inconsistent with the current data at more than $1\sigma$ level
\cite{star,lu09,huang,zhangxm}. Furthermore, it was suggested that
the cosmic acceleration is slowing down from $z\sim 0.3$ in
\cite{star}. The analysis in \cite{star} is based on the commonly
used Chevallier-Polarski-Linder (CPL) model \cite{cpl1,linder03}
with the Constitution SnIa \cite{consta} and the baryon acoustic
oscillation (BAO) distance ratio data \cite{sdss6}, the result is
somewhat model dependent. In \cite{lu09}, the authors applied the
Union SnIa data \cite{union}, together with the BAO $A$ parameter
\cite{bao} and gamma-ray bursts data to the piecewise constant
parametrization of the equation of state parameter of dark energy.
In this analysis, the equation of state parameter $w$ of dark energy
is a constant in a redshift bin, and three redshift bins with
boundaries at $z=0.2$, $0.5$ and $1.0$ were chosen. They also
analyzed the Constitution SnIa and the BAO distance ratio data. In
\cite{huang}, the authors considered two redshift bins by using the
Constitution SnIa data, and they found that dark energy suddenly
emerged at redshift $z\sim 0.3$. The results obtained in
\cite{zhangxm} were based on the analysis of the Constitution SnIa,
the full Wilkinson microwave anisotropy probe 5 year (WMAP5)
\cite{wmap5} and the Sloan digital sky survey (SDSS) data. For the
redshift range $0\le z\le 1$, four evenly spaced redshift bins were
chosen. However, choosing the same four evenly spaced redshift bins,
the authors in \cite{corray9} found no evidence for dark energy
dynamics by using the Constitution SnIa, WMAP5, BAO \cite{sdss7},
the integrated Sachs-Wolfe effect, galaxy clustering and weak
lensing data. Different data sets and analysis may give different
results. In this paper, we apply the piecewise constant
parametrization of the equation of state parameter $w(z)$ to do a
more careful model independent analysis. We choose four redshift
bins by requiring $N\Delta z\sim 30$ in each bin, and we use the
Constitution SnIa \cite{consta}, the BAO \cite{sdss6}, the derived
WMAP5 \cite{wmap5} and the Hubble parameter $H(z)$ data
\cite{hz1,hz2,riess09}.

This paper is organized as follows. In section II, we first review
the analysis by using the CPL model in \cite{star}, and find that
their result heavily depends on the choice of BAO data. By using the
BAO distance ratio data, the best fit value of $\Omega_{m0}=0.45^{+0.07}_{-0.11}$,
which is not consistent with other observational result. On
the other hand, the best fit value of $\Omega_{m0}=0.29^{+0.05}_{-0.04}$ if
the BAO $A$ parameter is used. Then we apply the SnIa, BAO, WMAP5 and $H(z)$ data to study
the property of dark energy by using the piecewise
constant parametrization of the equation of state parameter $w(z)$, in section III. We conclude
the paper in section IV.

\section{Observational constraints on CPL parametrization}

\begin{figure}[htp]
\centering
\includegraphics[height=0.6\textheight,width=\textwidth]{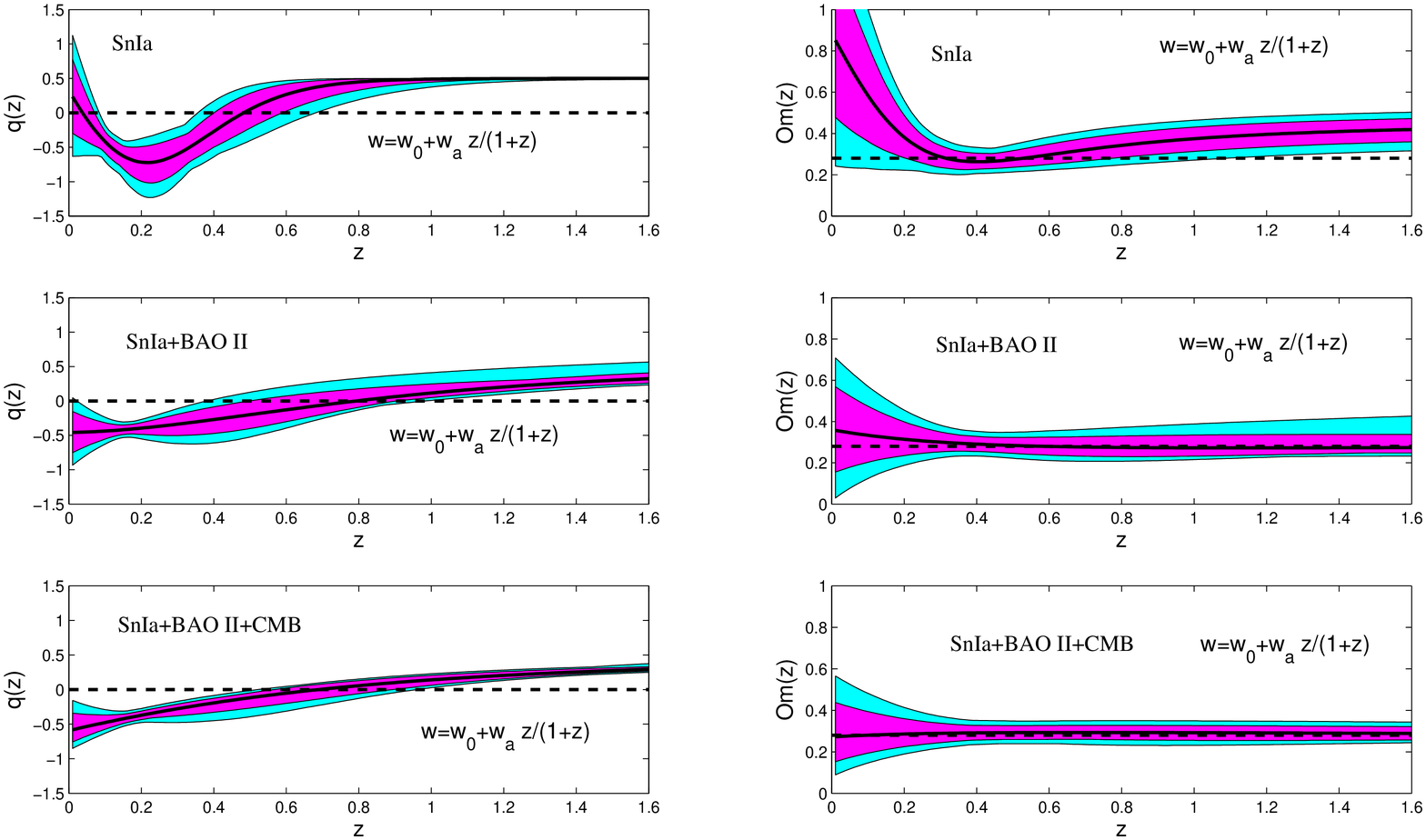}
\caption{Reconstructed $q(z)$ and $Om(z)$ from SnIa data, SnIa+BAO II data
and SnIa+BAO II+CMB data using the CPL ansatz. The
center solid lines are plotted with the best fit values, and the
shadows denote the $1\sigma$ and $2\sigma$ limits. The spatially
flat $\Lambda$CDM model corresponds to a horizonal dashed line with
$Om(z)=0.28$ in the right panels.} \label{qOm}
\end{figure}

To study the dynamical property of dark energy by observational
data, one usually parameterizes the equation of state parameter
$w(z)$. Following \cite{star}, we first study the CPL
parametrization
\begin{equation}
\label{cplwz}
w(z)=w_0+\frac{w_a \, z}{1+z}.
\end{equation}
The dimensionless Hubble parameter for a flat universe is
\begin{equation}
\label{cplez}
E^2(z)=\frac{H^2(z)}{H^2_0}=\Omega_{m0}(1+z)^3+(1-\Omega_{m0})(1+z)^{3(1+w_0+w_a)}\exp(-3w_a
z/(1+z)).
\end{equation}
In this model, we have three parameters $\Omega_{m0}$, $w_0$ and $w_a$, let us denote them
as $\bm{p}=(\Omega_{m0}, w_0, w_a)$. We first use the Constitution compilation of 397 SnIa data \cite{consta}
to constrain the model parameters $\bm{p}$. The Constitution sample adds 185 CfA3 SnIa
data to the Union sample \cite{union}. The addition of CfA3 sample
increases the number of nearby SnIa by a factor of roughly
$2.6-2.9$ and reduces the statistical uncertainties. The Union
compilation has 57 nearby SnIa and 250 high-$z$ SnIa. It includes
the Supernova Legacy Survey \cite{astier} and the ESSENCE Survey
\cite{riess,essence}, the older observed SnIa data, and the
extended data set of distant SnIa observed with the Hubble space
telescope. To fit the SnIa data, we define
\begin{equation}
\label{chi}
\chi^2_{sn}(\bm{p},H_0^n)=\sum_{i=1}^{397}\frac{[\mu_{obs}(z_i)-\mu(z_i,\bm{p},H_0^n)]^2}{\sigma^2_i},
\end{equation}
where the extinction-corrected distance modulus $\mu(z)$ is the difference between the apparent magnitude $m(z)$
and the absolute magnitude $M$ of a supernova at redshift $z$,
$$\mu(z,\bm{p},H_0^n)=m(z)-M=25-5\log_{10}H_0^n+5\log_{10}[D_L(z)/{\rm Mpc}],$$
the absolute magnitude $M$ applies equally to all magnitude measurement, and its effect
is manifested by the nuisance parameter $H_0^n$; $\sigma_i$ is the total
uncertainty which includes the intrinsic uncertainty of 0.138 mag for each CfA3 SnIa, the peculiar
velocity uncertainty of $400$km/s, and the redshift uncertainty \cite{consta}; and the
Hubble constant free luminosity distance $D_L(z)=H_0 d_L(z)$ is
\begin{equation}
\label{lum}
D_L(z,\bm{p})=H_0d_L(z)=\frac{1+z}{\sqrt{|\Omega_{k}|}} {\rm
sinn}\left[\sqrt{|\Omega_{k}|}\int_0^z
\frac{dx}{E(x,\bm{p})}\right],
\end{equation}
where
\begin{equation}
\frac{{\rm sinn}(\sqrt{|\Omega_k|}x)}{\sqrt{|\Omega_k|}}=\begin{cases}
\sin(\sqrt{|\Omega_k|}x)/\sqrt{|\Omega_k|},& {\rm if}\ \Omega_k<0,\\
x, & {\rm if}\  \Omega_k=0, \\
\sinh(\sqrt{|\Omega_k|}x)/\sqrt{|\Omega_k|}, & {\rm if}\  \Omega_k>0.
\end{cases}
\end{equation}
The nuisance parameter $H_0^n$ is marginalized over with a flat prior
when we apply the SnIa data. For the details of the marginalization method, see \cite{goliath,gong08}.
Then we add the BAO parameter $A=0.469(0.96/0.98)^{-0.35}\pm 0.017$ \cite{bao} into the SnIa data to determine the parameters $\bm{p}$.
So we have
$\chi^2(\bm{p})=[A-0.469(0.96/0.98)^{-0.35}]^2/0.017^2+\chi^2_{sn}$. The BAO
parameter $A$ in a spatially flat universe is defined as
\begin{equation}
\label{paraa}
A(\bm{p})=\sqrt{\Omega_{m0}}\,\frac{H_0 D_V(z_{bao},\bm{p},H_0)}{z_{bao}}
=\frac{\sqrt{\Omega_{m0}}}{z_{bao}}\left[\frac{z_{bao}}{E(z_{bao})}\left(\int^{z_{bao}}_0\frac{dz}{E(z)}\right)^2\right]^{1/3},
\end{equation}
where $z_{bao}=0.35$, and the effective distance
\begin{equation}
\label{dvdef}
D_V(z,\bm{p},H_0)=\frac{1}{H_0}\left[\frac{D_L^2(z)}{(1+z)^2}\frac{z}{E(z)}\right]^{1/3}.
\end{equation}
Finally we add the shift parameter $R$ with which the $l$-space positions of the acoustic peaks in the angular power spectrum shift,
to the combined SnIa and BAO $A$ data.
The shift parameter
\begin{equation}
\label{shift}
R(\bm{p})=\sqrt{\Omega_{m0}}\int_0^{z_{ls}}\frac{dz}{E(z)}=1.710\pm 0.019,
\end{equation}
where the last scattering surface redshift $z_{ls}=1090.0$. So now we minimize
$$\chi^2(\bm{p})=\frac{(R-1.71)^2}{0.019^2}+\frac{[A-0.469(0.96/0.98)^{-0.35}]^2}{0.017^2}+\chi^2_{sn}.$$
Fitting the Constitution SnIa data,  the BAO distance
ratio $D_{V}(z=0.35)/D_{V}(z=0.20)=1.736\pm0.065$ (hereafter BAO I)
\cite{sdss7} and the shift parameter $R$ to the
CPL model, the authors in \cite{star} conclude that the functional
form of the CPL ansatz is unable to fit the data simultaneously at
low and high redshifts. In this section, we replace the BAO distance ratio data (BAO I) by the BAO
parameter $A$ (hereafter BAO II) to fit the CPL model. By fitting the SnIa, SnIa+BAO I (II), and the SnIa+BAO I (II)+$R$ data
to the CPL model, we obtain the joint constraints on the parameters $\Omega_{m0}$, $w_0$ and $w_a$ in the CPL model.
The values of $\Omega_{m0}$ and
$\chi^2$ constrained from different data sets are shown in Table I. For the purpose of comparing the results,
we only show the constraint on $\Omega_{m0}$ in Table I.
For the SnIa+BAO I data, the joint $1\sigma$ constraints are $\Omega_{m0}=0.45^{+0.07}_{-0.11}$,
$w_0=-0.13^{+1.26}_{-0.95}$ and $w_a=-12.2^{+10.3}_{-15.3}$.
Fitting the SnIa+BAO II data to the CPL model, the joint $1\sigma$ constraints
are $\Omega_{m0}=0.29^{+0.05}_{-0.04}$,
$w_0=-0.90^{+0.46}_{-0.37}$ and $w_a=-0.6^{+2.5}_{-3.5}$. Comparing these results,
we find that $w_0$ and $w_a$ are consistent with each other at $1\sigma$ level,
but $\Omega_{m0}$ is barely consistent with each other at $1\sigma$ level.
For the SnIa data or the SnIa+BAO I data, the best fit value of $\Omega_{m0}$ is much larger
than the value $\Omega_{m0}\sim 0.3$ obtained from other
observational constraint, and that makes the result inconsistent
with the $\Lambda$CDM model at more than $1\sigma$ level. The BAO II data or the WMAP5 data
lowers the value of $\Omega_{m0}$, and therefore makes the result
consistent with the $\Lambda$CDM model at $1\sigma$ level.
\begin{table}[htp]
\begin{tabular}{|l|c|c|}
\hline
Data & CPL model & $\Lambda$CDM model \\\hline
SnIa & $\Omega_{m0}=0.45^{+0.07}_{-0.13}$, \ $\chi^2=462.07$ &  $\Omega_{m0}=0.29\pm 0.02$, \ $\chi^2=466.32$ \\
SnIa + BAO I & $\Omega_{m0}=0.45^{+0.07}_{-0.11}$, \  $\chi^2=462.44$ & $\Omega_{m0}=0.29\pm 0.02$, \ $\chi^2=467.61$ \\
SnIa + BAO II & $\Omega_{m0}=0.29^{+0.05}_{-0.04}$,\ $\chi^2=466.18$ & $\Omega_{m0}=0.28\pm 0.02$, \ $\chi^2=466.42$ \\
SnIa + BAO I + $R$ & $\Omega_{m0}=0.26^{+0.08}_{-0.04}$,\ $\chi^2=467.74$ & $\Omega_{m0}=0.27\pm 0.02$, \ $\chi^2=469.49$ \\
SnIa + BAO II + $R$ & $\Omega_{m0}=0.27\pm0.03$,\ $\chi^2=466.81$ & $\Omega_{m0}=0.27\pm 0.01$, \ $\chi^2=468.42$ \\
 \hline
\end{tabular}
\caption{The 1$\sigma$ error estimate of $\Omega_{m0}$ for the CPL model and $\Lambda$CDM model.}
\end{table}
To compare our results with those in \cite{star}, we also show the evolutions of $q(z)$ and $Om(z)$
constrained from SnIa, SnIa+BAO II and SnIa+BAO II+$R$
in Fig. \ref{qOm}. The deceleration parameter
\begin{equation}
\label{qzeq}
q(z)=\frac{\Omega_{m0}(1+z)^3+[1+3w(z)](1-\Omega_{m0})(1+z)^{3(1+w_0+w_a)}\exp(-3w_a
z/(1+z))}{2E^2(z)},
\end{equation}
and $Om(z)$ is defined as
\begin{equation}
\label{omzeq}
Om(z)=\frac{E^2(z)-1}{(1+z)^3-1}.
\end{equation}
The evolution of $q(z)$ gives us the information about how fast the Universe expands.
The sign of $q(z)$ shows whether the expansion is accelerating or decelerating, $q(z)>0$ means deceleration.
For a $\Lambda$CDM model, $Om(z)=\Omega_{m0}$ is a constant. Also at low redshift, larger $Om(z)$ means larger $w$ of dark energy.

Comparing the evolutions of $q(z)$ and $Om(z)$ shown in our Fig. \ref{qOm}
with those in the Fig. 2 of \cite{star}, we
find that the value of $\Omega_{m0}$ constrained from the sample is the main reason of
different evolutions of $q(z)$ and $Om(z)$. Because of the relative large $\Omega_{m0}$ obtained from SnIa + BAO I data,
the result is inconsistent with $\Lambda$CDM model at more than $1\sigma$ level. This can be easily
understood by expanding $Om(z)$ for the CPL model at low redshift,
$Om(z)\approx 1+w_0(1-\Omega_{m0})$. Even if we get $w_0$ close to $-1$, the large value
of $\Omega_{m0}$ will make the CPL model inconsistent with $\Lambda$CDM model. However, if we impose a prior on
$\Omega_{m0}$, for example, $\Omega_{m0}=0.28\pm 0.04$, then the
result by fitting the SnIa data or SnIa+BAO I data to the CPL
model, is consistent with the $\Lambda$CDM model.

\section{Observational Constraints on piecewise constant parametrization}

Although the CPL parametrization provides a useful tool to study the
dynamical property of dark energy, the particular form of $w(z)$ may
impose a strong prior. Note that in a small enough redshift region,
$w(z)$ is approximately a constant, so we may divide the redshift
into several bins, and parameterize $w(z)$ as a constant in a
particular redshift bin, this is the piecewise constant
parametrization of $w(z)$. If we have enough data, the redshift range
in each bin can be small enough, and the piecewise constant
parametrization gives the true $w(z)$. In other words, the piecewise
constant parametrization is a model independent method. In practice,
the number of redshift bins is finite, and the piecewise constant
parametrization of $w$ is an approximation of the true $w(z)$, and
it provides very useful information about the dynamical behavior of
dark energy. In this section, we use observational data to fit the piecewise constant
parametrization of $w(z)$. For the binning of the Constitution SnIa data \cite{consta},
we apply the uniform, unbiased binning method \cite{riess}.
We group the data into four bins so that the number of SnIa in each bin times the width of each bin is around 30, i.e.,
$N\Delta z\sim 30$, $N$ is the number of SnIa in each bin, $\Delta z$ is the width of each bin.
The choice of $N\Delta z\sim 30$ for the Constitution SnIa data results in four bins.
The boundaries of the four bins are $z_1=0.22$, $z_2=0.53$, $z_3=0.85$, $z_4=1.8$ and $z_5$ extends beyond 1089.
For the redshift in the range $z_{i-1}<z<z_i$, the equation of state parameter is a constant, $w(z)=w_i$. For convenience, we
choose $z_0=0$. Due to the lack of observational data in the redshift range $z=1.8-1089$,
$w(z)$ is largely unconstrained in this redshift range. For simplicity,
we assume that $w(z>1.8)=-1$. For a flat universe, the Friedmann equation becomes
\begin{equation}
\label{wcez}
E^2(z)=\Omega_{m0}(1+z)^3+(1-\Omega_{m0})(1+z)^{3(1+w_n)}\prod_{i=1}^{n}(1+z_{i-1})^{3(w_{i-1}-w_i)},\quad
z_{n-1}<z<z_n.
\end{equation}
In this model, there are five free parameters $\Omega_{m0}$, $w_1$, $w_2$, $w_3$ and $w_4$, let us denote them as
$\bm{\theta}=(\Omega_{m0},w_1,w_2,w_3,w_4)$.

In general, the equation of state parameters $w_i$ in different bins
are correlated and their errors depend upon each other. We follow
Huterer and Cooray \cite{par5} to transform the covariance matrix of
$w_i$ to decorrelate the error estimate. Explicitly, the
transformation is
\begin{equation}
\label{decorr}
{\mathcal W}_i=\sum_j T_{ij}w_j,
\end{equation}
where the transformation matrix $T=V^T\Lambda^{-1/2}V$, the orthogonal matrix
$V$ diagonalizes the covariance matrix $C$ of $w_i$ and $\Lambda$ is the diagonalized matrix of $C$.
For a given $i$, $T_{ij}$ can be thought of as weights for each $w_j$ in the transformation from
$w_i$ to ${\mathcal W}_i$. We are free to rescale each ${\mathcal W}_i$ without changing the
diagonality of the correlation matrix, so we then multiply both sides of the equation above by an
amount such that the sum of the weights $\sum_j T_{ij}$ is equal to one. This allows for easy
interpretation of the weights as a kind of discretized window function. Now the transformation matrix
element is $T_{ij}/\sum_k T_{ik}$ and the covariance matrix of the uncorrelated parameters is not the
identity matrix. The $i$-th diagonal matrix element becomes $(\sum_j T_{ij})^{-2}$. In other words,
the error of the uncorrelated parameters ${\mathcal W}_i$ is $\sigma_i=1/\sum_j T_{ij}$.

The likelihood for the parameters $\bm{\theta}$ in the
model and the nuisance parameters is
computed using a Monte Carlo Markov Chain (MCMC). To observe the effect of the nuisance parameter
$H_0^n$ in the SnIa data, we take two different approaches. In the first approach, we analytically
marginalize $H_0^n$ by using a flat prior \cite{gong08}. In the second approach, we take $H_0^n=H_0$ as a free parameter
in the MCMC code. The MCMC method
randomly chooses values for the above parameters $\bm{\theta}$, evaluates $\chi^2$
and determines whether to accept or reject the set of parameters $\bm{\theta}$
using the Metropolis-Hastings algorithm. The set of parameters that
are accepted to the chain forms a new starting point for the next
process, and the process is repeated for a sufficient number of
steps until the required convergence is reached. Our MCMC code is
based on the publicly available package COSMOMC \cite{cosmomc}.
We give both marginalized and likelihood limits of
the uncorrelated parameters ${\mathcal W}_i$. The likelihood limit defines the region of parameter space
enclosing a fraction $f$ of the points with the highest likelihood as the $N$-dimensional confidence region,
where $f$ defines the confidence limit \cite{cosmomc}. The likelihood limit is very useful
to assess the consistency with new data or theories.
For comparison, we also fit the result by running the publicly available package WZBINNED \cite{wzbin}
and the results are consistent with those by using the marginalized method.

\begin{figure}[htp]
\centering
\includegraphics[width=12cm]{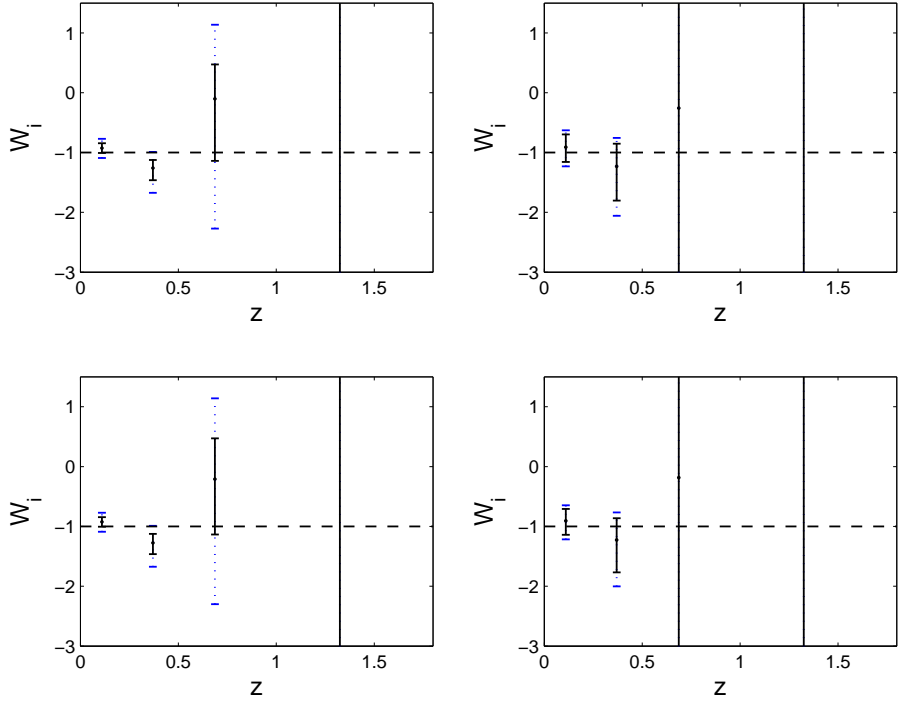}
\caption{The $1\sigma$ and $2\sigma$ estimates of
the four uncorrelated parameters ${\mathcal W}_i$ using the SnIa+BAO II data.
The parameter ${\mathcal W}_4$ is not well constrained. The error bars show $1\sigma$
and $2\sigma$ uncertainties with black solid lines and blue dotted lines, respectively. The top panels show the results
by treating the nuisance parameter $H_0^n$ as a free parameter, and the bottom panels show the
results with $H_0^n$ being analytically marginalized. The results in the left panels are the marginalized limits,
while the results in the right panels are the likelihood limits.}
\label{snwunc}
\end{figure}

\begin{figure}[htp]
\centering
\includegraphics[width=12cm]{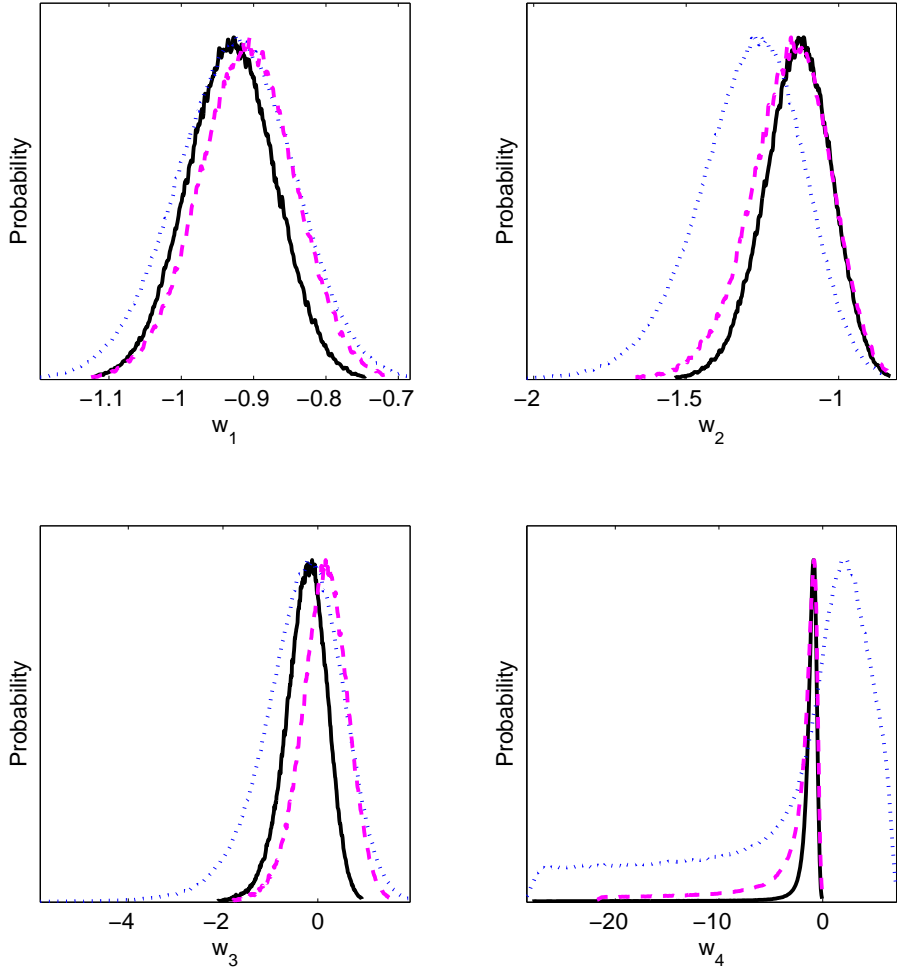}
\caption{The marginalized probabilities of the four equation of state parameters ${\mathcal W}_i$.
The dotted lines are the results using
SnIa+BAO II data only, the dashed lines are the results using
the combined SnIa, BAO III and WMAP5 data, and the solid lines are the results using SnIa+BAO III+WMAP5+$H(z)$ data.
The nuisance parameter $H_0^n$ in the SnIa is analytically marginalized over.}
\label{mprob}
\end{figure}

We first fit the parameters $\bm{\theta}$ in the model by using the combined SnIa+BAO II data, i.e.,
we calculate $\chi^2(\bm{\theta})=[A(\bm{\theta})-0.469(0.96/0.98)^{-0.35}]^2/0.017^2+\chi^2_{sn}(\bm{\theta},H^n_0)$.
By fitting the piecewise constant model to the data, we get $\chi^2=459.2$,
the marginalized $1\sigma$ estimate $\Omega_{m0}=0.285_{-0.011}^{+0.034}$, and the uncorrelated binned
estimates of the equation of state parameters ${\mathcal W}_i$ are
shown in Fig. \ref{snwunc}. We see that the results marginalizing
over $H_0^n$ analytically are the same as those with $H_0$
marginalized numerically. In other words, we can trust the result
using the method of analytically marginalizing over $H_0^n$ with a
flat prior. From Fig. \ref{snwunc}, it is clear that the likelihood
limits are larger than the marginalized limits, especially for the
parameters ${\mathcal W}_3$ and  ${\mathcal W}_4$. If we take the
likelihood limits, then the cosmological constant is consistent with
the data even at $1\sigma$ level. However, the cosmological constant
is not consistent with the data at $1\sigma$ level if we take the
marginalized limits. Although the likelihood limits are useful
information for the full MCMC sample, in the following, we quote the
results obtained with marginalized limits. If we fit the flat $\Lambda$CDM model to the combined SnIa+BAO II
data, we get $\chi^2=466.4$ and $\Omega_{m0}=0.286\pm 0.017$. If we
fit the flat CPL model to the data, we get $\chi^2=466.2$, and the marginalized $1\sigma$ constraints are
$\Omega_{m0}=0.288^{+0.041}_{-0.016}$, $w_0=-0.90^{+0.30}_{-0.16}$,
and $w_a=-0.6^{+1.1}_{-2.3}$. In Fig. \ref{mprob}, we
show the marginalized probabilities of the parameters ${\mathcal W}_i$.

Now we add the BAO and WMAP5 data to the SnIa data. The parameters $\bm{\theta}$ in the models are
determined by minimizing
$\chi^2(\bm{\theta},\Omega_b h^2, h)=\chi^2_{sn}+\chi^2_{bao}+\chi^2_{cmb}$, here $h=H_0/100$.
To use the BAO measurement (hereafter BAO III) from the SDSS data, we define
\cite{sdss6}
\begin{equation}
\label{baochi2}
\chi^2_{bao}(\bm{\theta},\Omega_b h^2,h)=\left(\frac{r_s(z_d)/D_V(z=0.2)-0.198}{0.0058}\right)^2
+\left(\frac{r_s(z_d)/D_V(z=0.35)-0.1094}{0.0033}\right)^2,
\end{equation}
where the redshift $z_d$ is fitted with the formulae \cite{hu98}
\begin{equation}
\label{zdfiteq}
z_d=\frac{1291(\Omega_{m0} h^2)^{0.251}}{1+0.659(\Omega_{m0} h^2)^{0.828}}[1+b_1(\Omega_b h^2)^{b_2}],
\end{equation}
\begin{equation}
\label{b1eq}
b_1=0.313(\Omega_{m0} h^2)^{-0.419}[1+0.607(\Omega_{m0} h^2)^{0.674}],\quad b_2=0.238(\Omega_{m0} h^2)^{0.223},
\end{equation}
and the comoving sound horizon is
\begin{equation}
\label{rshordef}
r_s(z)=\int_z^\infty \frac{dx}{c_s(x)E(x)},
\end{equation}
where the sound speed $c_s(z)=1/\sqrt{3[1+\bar{R_b}/(1+z)}]$,
and $\bar{R_b}=315000\Omega_b h^2(2.726/2.7)^{-4}$.

To implement the WMAP5 data, we need to add three fitting parameters
$R$, $l_a$ and $z_*$, so $\chi^2_{cmb}(\bm{\theta},\Omega_b h^2, h)=\Delta x_i {\rm
Cov}^{-1}(x_i,x_j)\Delta x_j$, where $x_i=(R,\ l_a,\ z_*)$ denote
the three parameters for WMAP5 data, $\Delta x_i=x_i-x_i^{obs}$ and
Cov$(x_i,x_j)$ is the covariance matrix for the three parameters
\cite{wmap5}. The acoustic scale $l_A$ is
\begin{equation}
\label{ladefeq}
l_A=\frac{\pi d_L(z_*)}{(1+z_*)r_s(z_*)},
\end{equation}
where the redshift $z_*$ is given by \cite{hu96}
\begin{equation}
\label{zstareq}
z_*=1048[1+0.00124(\Omega_b h^2)^{-0.738}][1+g_1(\Omega_{m0} h^2)^{g_2}]=1090.04\pm 0.93,
\end{equation}
\begin{equation}
g_1=\frac{0.0783(\Omega_b h^2)^{-0.238}}{1+39.5(\Omega_b h^2)^{0.763}},\quad
g_2=\frac{0.560}{1+21.1(\Omega_b h^2)^{1.81}}.
\end{equation}
The shift parameter \cite{wmap5}
\begin{equation}
\label{shift1}
R(\bm{\theta},\Omega_b h^2,h)=\sqrt{\Omega_{m0}}\int_0^{z_*}\frac{dz}{E(z)}=1.710\pm 0.019,
\end{equation}

\begin{figure}[htp]
\centering
\includegraphics[width=12cm]{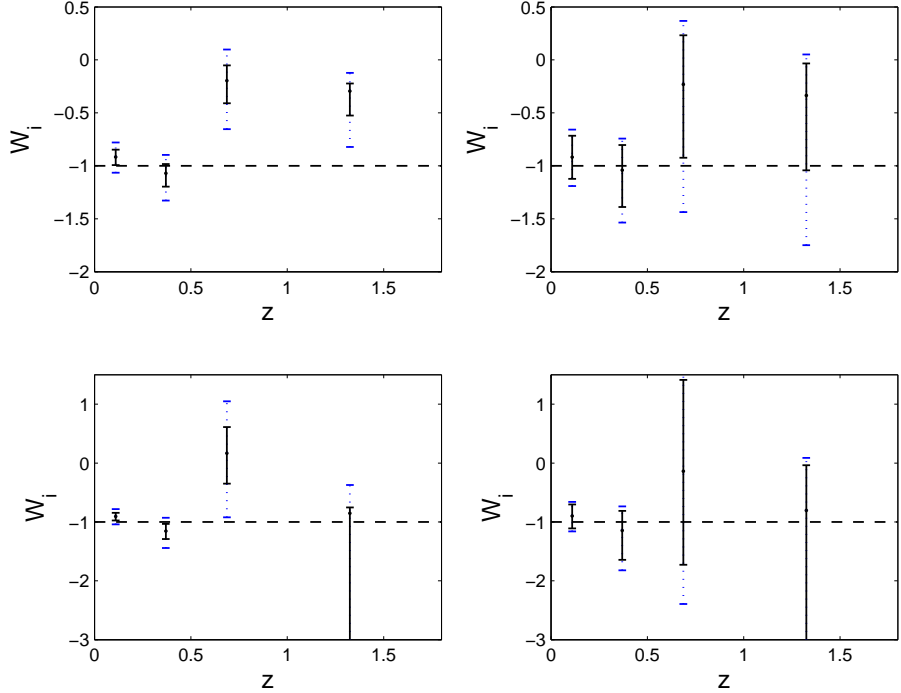}
\caption{Same as Fig. \ref{snwunc} except that we use SnIa + BAO III + WMAP5 data here.}
\label{obswunc}
\end{figure}

When we add the BAO III and WMAP5 data, we add two more parameters $\Omega_b h^2$ and the
Hubble constant $H_0=100 h$. Because the normalization of the luminosity distance-redshift relation
is unknown, the nuisance parameter $H_0^n$ in the SnIa data is not the observed Hubble constant,
and it is different from that in the BAO III and WMAP5 data. By fitting the data to the model, we
find that the nuisance parameter $H_0^n$ is around 65 km/s/Mpc in the Constitution data, while
the Hubble constant $H_0$ is around 72 km/s/Mpc. Therefore, we should treat the nuisance parameter $H_0^n$
in the SnIa data differently from the Hubble constant $H_0$ in other observational data, and we
should include both $H_0^n$ and $H_0$ ($h$) in the data fitting. We
analytically marginalize over the nuisance parameter $H_0^n$ in the SnIa data as explained in \cite{gong08}.
If we treat the nuisance parameter $H_0^n$ as the Hubble constant, we get $\chi^2=465.9$, and the uncorrelated
estimates of ${\mathcal W}_i$ are shown in the top panels of Fig. \ref{obswunc}. If the nuisance parameter
$H_0^n$ in the SnIa data is marginalized analytically, we get $\chi^2=462.3$, $\Omega_{m0}=0.283^{+0.020}_{-0.011}$,
and the uncorrelated estimates
of ${\mathcal W}_i$ are shown in the bottom panels of Fig. \ref{obswunc}. By adding the BAO III and WMAP5 data, we add
4 more data points, and $\chi^2$ increases 3.1.
With the help of the BAO III and WMAP5 data,
the constraint on the equation of state parameters ${\mathcal W}_i$ is improved, especially for
${\mathcal W}_3$ and ${\mathcal W}_4$. This point is also clear from the marginalized probabilities shown in
Fig. \ref{mprob}.
From Fig. \ref{obswunc}, we see that the results are different if the nuisance parameter $H_0^n$ is treated
differently. As explained above and in \cite{gong08}, we should treat the nuisance parameter $H_0^n$ in the SnIa data
differently, so we analytically marginalize over $H_0^n$ with a flat prior when fitting the SnIa data.
Again, we see that the marginalized limits give tighter constraints on the parameters than the likelihood limits do,
so we quote the result in the bottom left panel as the fitting result.

Finally, we add the data of the Hubble parameter $H(z)$ at
nine different redshifts from the differential ages of passively
evolving galaxies obtained in \cite{hz1} and the three more recent data
$H(z=0.24)=79.69\pm 2.32$, $H(z=0.34)=83.8\pm 2.96$, and $H(z=0.43)=86.45\pm 3.27$ by
taking the BAO scale as a standard ruler in the radial direction \cite{hz2}. To
use these 12 $H(z)$ data, we define
\begin{equation}
\label{hzchi}
\chi^2_h(\bm{\theta},h)=\sum_{i=1}^{12}\frac{[H_{obs}(z_i)-H(z_i)]^2}{\sigma_{hi}^2},
\end{equation}
where $\sigma_{hi}$ is the $1\sigma$ uncertainty in the $H(z)$ data.
We also add the prior $H_0=74.2\pm 3.6$ km/s/Mpc determined from the observations with the Hubble Space Telescope by Riess {\it
et al.} \cite{riess09}. Now we have
$\chi^2(\bm{\theta},\Omega_b h^2,h)=\chi^2_{sn}+\chi^2_{bao}+\chi^2_{cmb}+\chi^2_h$.

\begin{figure}[htp]
\centering
\includegraphics[width=12cm]{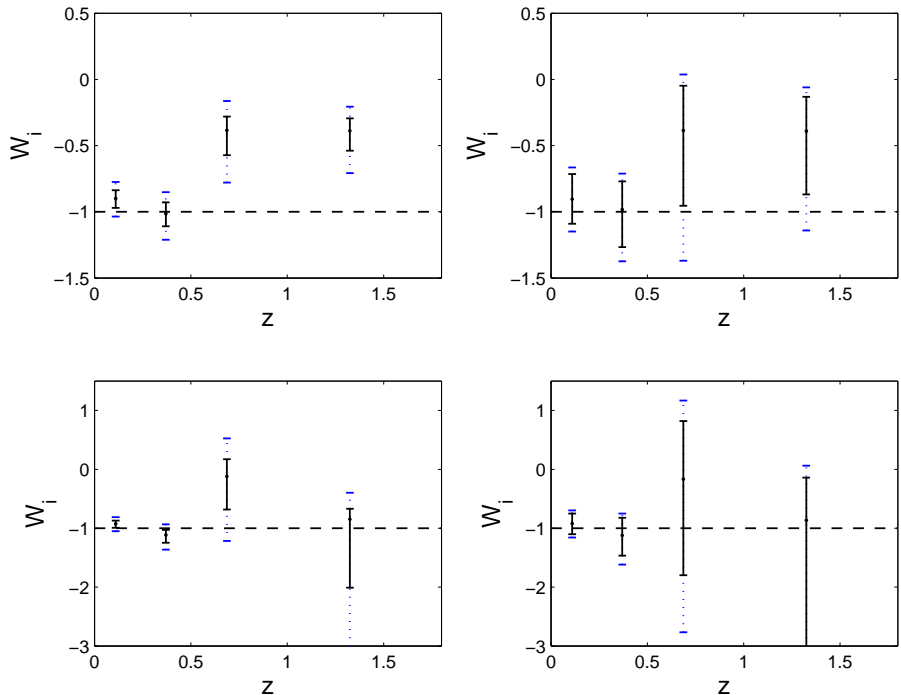}
\caption{Same as Fig. \ref{snwunc}, except that we use the combined SnIa, BAO III, WMAP5 and $H(z)$ data here.}
\label{obswhz}
\end{figure}

If we treat the nuisance parameter $H_0^n$ as the Hubble constant, we get $\chi^2=500.5$, and the uncorrelated
estimates of ${\mathcal W}_i$ are shown in the top panels of Fig. \ref{obswhz}. If the nuisance parameter
$H_0^n$ in the SnIa data is marginalized analytically, we get $\chi^2=476.8$, $\Omega_{m0}=0.271^{+0.021}_{-0.006}$,
 and the uncorrelated estimates
of ${\mathcal W}_i$ are shown in the bottom panels of Fig. \ref{obswhz}. By adding the $H(z)$ data, we add 13 data points, and
$\chi^2$ increases 14.5.
As explained above, due to the difference
between the nuisance parameter $H_0^n$ in the SnIa data and the Hubble constant $H_0$, we get much larger value
of $\chi^2$ if we treat the nuisance parameter $H_0^n$ in the SnIa data as the Hubble constant,
so we should use the results in the bottom left panel.
If we fit the data to the flat $\Lambda$CDM model, we get $\chi^2=483.0$ and $\Omega_{m0}=0.272\pm 0.011$.
If we fit the data to the flat CPL model, we get $\chi^2=482.5$, $\Omega_{m0}=0.269^{+0.017}_{-0.008}$,
$w_0=-0.97^{+0.12}_{-0.07}$, and $w_a=0.03^{+0.26}_{-0.75}$.
These results suggest that the CPL model is consistent with $\Lambda$CDM model at $1\sigma$ level,
and $\Lambda$CDM model is consistent with the piecewise constant parametrization at $2\sigma$ level.

The marginalized
probabilities of the parameters ${\mathcal W}_i$ are shown in Fig. \ref{mprob}.
From Figs. \ref{mprob} and \ref{obswhz}, we see that the addition of the $H(z)$ data improves the constraint
on ${\mathcal W}_4$ by 70\%. This suggests that the equation of state parameter at high redshift will be better constrained
with high quality data of $H(z)$ in the future.

\section{Conclusions}

Fitting the combined SnIa and the BAO distance ratio (BAO I) data to the CPL model, we
find that $\chi^2=462.44$, and the joint $1\sigma$ constraints $\Omega_{m0}=0.45^{+0.07}_{-0.11}$,
$w_0=-0.13^{+1.26}_{-0.95}$ and $w_a=-12.2^{+10.3}_{-15.3}$.
Fitting the combined SnIa and the BAO $A$ (BAO II) data to the CPL model, we
find that $\chi^2=466.18$, and the joint $1\sigma$ constraints $\Omega_{m0}=0.29^{+0.05}_{-0.04}$,
$w_0=-0.90^{+0.46}_{-0.37}$ and $w_a=-0.6^{+2.5}_{-3.5}$. While the results obtained with
the BAO I data are not consistent with the $\Lambda$CDM model at more than $1\sigma$ level \cite{star}, the results obtained with
the BAO II data are consistent with the $\Lambda$CDM model at $1\sigma$ level (see Fig. \ref{qOm}), and the constraints are tighter.
So different BAO data give different results. The inconsistency lies mainly on the larger
value of $\Omega_{m0}$ obtained with the BAO I data.

To constrain the property of dark energy using the observational data, we need to apply model independent
method. The piecewise constant parametrization of the equation of state parameter $w(z)$ of dark energy
is somewhat model independent, we used the current observational data to study the property of dark energy
with this model independent parametrization. Since the normalization of the luminosity
distance-redshift relation is arbitrary, the nuisance parameter $H_0^n$ in the SnIa data is also arbitrary,
and different from the observed Hubble constant $H_0$. We should treat it differently from the
Hubble constant in other data, and we should include both $H_0^n$ and $H_0$ ($h$) in the data fitting.
Otherwise, we may get wrong conclusions. If we treat the nuisance parameter $H_0^n$ in the SnIa data as the observed Hubble constant,
then we may conclude that the flat $\Lambda$CDM model is incompatible with the combined SnIa, BAO III and WMAP5
data or the combined SnIa, BAO III, WMAP5 and $H(z)$ data. However,
by marginalizing over the nuisance parameter $H_0^n$ analytically in computing $\chi^2_{sn}$, the flat $\Lambda$CDM
model is consistent with the current observation at $2\sigma$ level.

By fitting the combined SnIa, BAO III, WMAP5 and $H(z)$ data to the flat $\Lambda$CDM model,
we get $\chi^2=483.0$ and $\Omega_{m0}=0.272\pm 0.011$.
If we fit the data to the flat CPL model, we get $\chi^2=482.5$, and the marginalized
$1\sigma$ constraints $\Omega_{m0}=0.269^{+0.017}_{-0.008}$,
$w_0=-0.97^{+0.12}_{-0.07}$, and $w_a=0.03^{+0.26}_{-0.75}$.
If we fit the model with piecewise constant parametrization to the data, we get
$\chi^2=476.8$, and the marginalized
$1\sigma$ constraints $\Omega_{m0}=0.271^{+0.021}_{-0.006}$, ${\mathcal W}_1=-0.93\pm 0.06$,
${\mathcal W}_2=-1.13^{+0.10}_{-0.12}$, ${\mathcal W}_3=-0.12^{+0.29}_{-0.56}$, and ${\mathcal W}_4=-0.85^{+0.18}_{-1.16}$.

By adding the $H(z)$ data to the SnIa, BAO III and WMAP5 data, the constraint on the parameters is greatly improved,
especially for the equation of state parameter at high redshift where the number of SnIa data is small.
The result suggests that the equation of state parameter at high redshift will be better constrained
with high quality data of $H(z)$ in the future.

\begin{acknowledgments}

YG and ZZ wish to acknowledge the hospitality of the KITPC under
their program ``Connecting Fundamental Theory with Cosmological
Observations'' during which this project was initiated. YG thanks A.
Cooray, D. Sarkar, A. Riess and B. Gold for the fruitful discussion
on the MCMC code. This work was partially supported by the Chinese
Academy of Sciences under grant No. KJCX3-SYW-N2, the National
Natural Science Foundation of China key project under grant
Nos. 10533010 and 10935013, and the Ministry of Science and
Technology of China national basic research Program (973 Program)
under grant Nos. 2007CB815401 and 2010CB833004.
ZZ acknowledges the supported by the National
Natural Science Foundation of China under the Distinguished Young Scholar Grant No. 10825313.
YG acknowledges the support by the Natural
Science Foundation Project of CQ CSTC under grant No. 2009BA4050.

\end{acknowledgments}

\end{document}